\documentclass[sigconf,screen]{acmart}
\usepackage{graphicx}
\usepackage{amsmath}

\usepackage{amssymb}
\usepackage{subfigure}
\usepackage{multirow}
\usepackage{balance}

\AtBeginDocument{%
	\providecommand\BibTeX{{%
			\normalfont B\kern-0.5em{\scshape i\kern-0.25em b}\kern-0.8em\TeX}}}

\copyrightyear{2021}
\acmYear{2021}
\setcopyright{acmcopyright}\acmConference[MM '21]{Proceedings of the 29th ACM
	International Conference on Multimedia}{October 20--24, 2021}{Virtual Event, China}
\acmBooktitle{Proceedings of the 29th ACM International Conference on Multimedia
	(MM '21), October 20--24, 2021, Virtual Event, China}
\acmPrice{15.00}
\acmDOI{10.1145/3474085.3475710}
\acmISBN{978-1-4503-8651-7/21/10}

\acmSubmissionID{3026}

\begin{document}
	\fancyhead{}
	\title{Recursive Fusion and Deformable Spatiotemporal Attention for Video Compression Artifact Reduction}
	
	
	
	\author{Minyi Zhao$^*$, Yi Xu$^*$, Shuigeng Zhou$^+$}
	\thanks{\textsuperscript{*}Both authors contributed equally to this research.}
	\thanks{\textsuperscript{+}Corresponding author.}
	\affiliation{
		\institution{Shanghai Key Lab of Intelligent Information Processing, and School of Computer Science, Fudan University}
		\city{Shanghai}
		\country{China}
	}
	\email{{zhaomy20,yxu17,sgzhou}@fudan.edu.cn}
	
	\begin{abstract}
		A number of deep learning based algorithms have been proposed to recover high-quality videos from low-quality compressed ones. Among them,
		some restore the missing details of each frame via exploring the spatiotemporal information of neighboring frames.
		However, these methods usually suffer from a narrow temporal scope, thus may miss some useful details from some frames outside the neighboring ones.
		In this paper, to boost artifact removal, on the one hand, we propose a \emph{Recursive Fusion} (RF) module to model the temporal dependency within a long temporal range.
		Specifically, RF utilizes both the current reference frames and the preceding hidden state to conduct better spatiotemporal compensation. 
		On the other hand, we design an efficient and effective \emph{Deformable Spatiotemporal Attention} (DSTA) module such that the model can pay more effort on restoring the artifact-rich areas like the boundary area of a moving object.
		Extensive experiments show that our method outperforms the existing ones on the MFQE~2.0 dataset in terms of both fidelity and perceptual effect. Code is available at \url{https://github.com/zhaominyiz/RFDA-PyTorch}.
	\end{abstract}


	
	
	\maketitle
	\begin{figure*}
		\begin{center}
			\includegraphics[width=\linewidth]{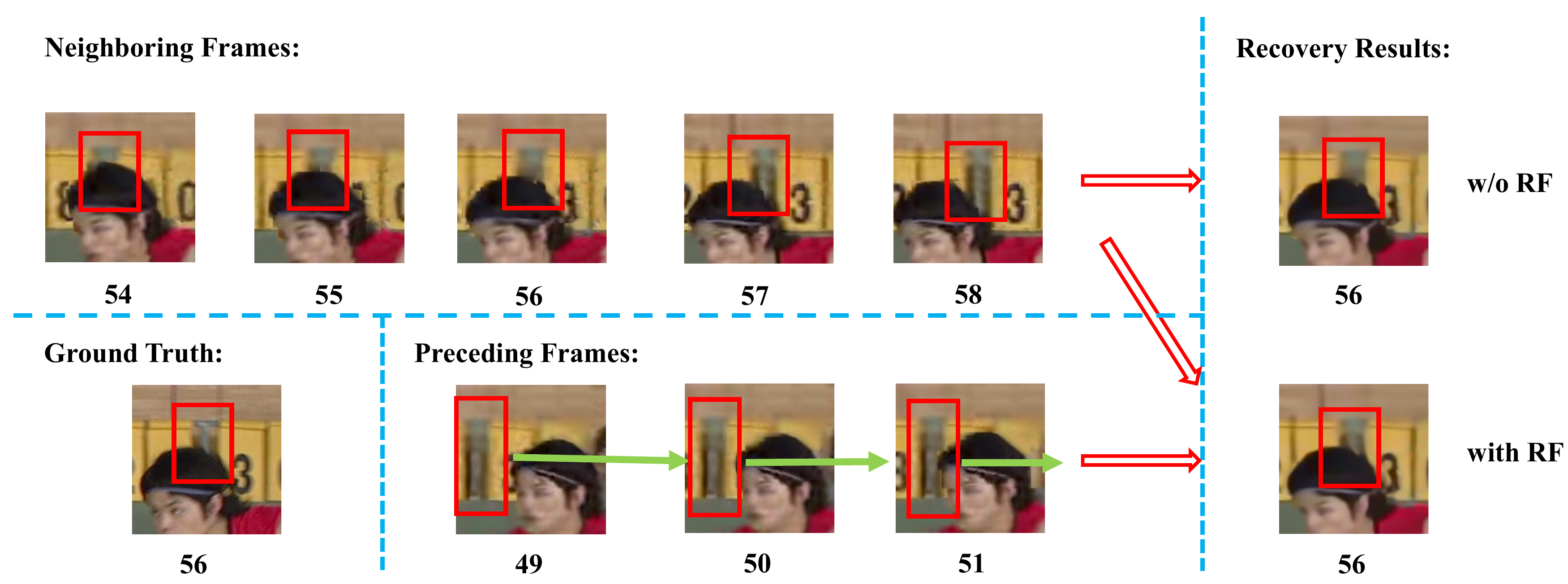}
		\end{center}
		\caption{An illustrative example of our proposed method. To recover the 56th frame, recent methods take from the 53rd to the 59th frames as reference, which cannot provide enough information to recover the area (just above the man's head in the red box) between the two scoreboards because in these neighboring frames this area is mostly occluded by the man. However, the proposed RF module can exploit extra~({\emph{e.g. }} the 49th, 50th and 51st) frames to restore this area better.}
		\label{fig-motivation}
	\end{figure*}
	
	\section{Introduction}
	Nowadays, lossy video compression algorithms (\emph{e.g.} H.264/AVC \cite{wiegand2003overview} and H.265/HEVC~\cite{sullivan2012overview}) are extensively applied for reducing the cost of video storage and transmission.
	As the compression rate increases, these algorithms reduce the bit-rate greatly but also introduce undesirable artifacts, which severely degrade the Quality of Experience (QoE).
	Besides, the artifacts in low-quality compressed videos 
	also harm the performance of down-stream video-oriented tasks (\emph{e.g.} action recognition and localization~\cite{zhu2020comprehensive,xia2020survey}, video summarization~\cite{apostolidis2021video, xu2021gif}).
	Accordingly, video compression artifact reduction, which aims to reduce the introduced artifacts and recover missed details for heavily compressed videos, becomes a hot topic in the multimedia field~\cite{2018Multi,2019MFQE,xu2019non,deng2020spatio,yang2021ntire, xu2021boosting}.
	
	Recently, many deep neural network based compression artifact removal works have emerged with significant performance improvement. These works roughly fall into three types: single-image  based~\cite{dong2015compression,zhang2017beyond,li2017efficient,chen2018dpw,guo2016building,taipersistent}, various video compression priors based~\cite{dai2017convolutional,jin2018quality,yang2017decoder}, and additional temporal information based~\cite{2018Multi,xue2019video,lu2018deep,yang2019quality,2019MFQE,xu2019non,lu2019deep,deng2020spatio,ding2021patch,xu2021boosting}, respectively. In detail, \cite{dong2015compression,zhang2017beyond,li2017efficient} are designed for JPEG quality enhancement. These methods can be adapted to videos by restoring each frame individually. \cite{dai2017convolutional,jin2018quality,yang2017decoder} consider the fact that I/P/B frames are compressed with different strategies and should be processed by different modules, they take a single frame as input, and ignore the temporal information of videos. In order to remedy this defect, \cite{2018Multi,2019MFQE} exploit two nearest high-quality peak-quality frames~(PQFs) as reference frames, \cite{lu2018deep,lu2019deep} utilize the deep Kalman filter network and capture spatiotemporal information from preceding frames, and \cite{xu2019non,deng2020spatio} employ non-local ConvLSTM and deformable convolution respectively to capture dependency among multiple neighboring frames. In summary, recent works employ either the preceding frames, nearby PQFs, or multiple adjacent frames as reference frames to exploit the spatiotemporal information among video frames. Although these methods have made great progress in this task, their performance is still limited by the narrow temporal scope, which makes them fail to fully exploit the spatiotemporal information in the preceding frames.
	
	To address the aforementioned issues, on the one hand, we propose a Recursive Fusion~(RF) module for video compression artifact reduction.
	Specifically, in order to leverage the relevant information in a large temporal scope, we develop a recursive fusion scheme and combine the preceding compensated feature with the current feature recursively with limited extra computational cost. Fig.~\ref{fig-motivation} shows an example that RF exploits the details from far away frames.
	As we can see, when the RF module is not used, we cannot restore the details in the 56th frame, while our method with the RF module can recover the details of the area between the two scoreboards successfully by exploiting the corresponding areas in the 49-51th frames. This is attributed to the enlarged temporal receptive field of the RF module.
	On the other hand, it is unreasonable to treat different areas of a frame equally when reconstructing its high-quality version. For example, the boundary area of a moving object usually suffers from severe distortions, thus should receive more attention in artifact reduction.
	Therefore, 
	we design an effective and efficient Deformable Spatiotemporal Attention mechanism~(DSTA) such that the model can pay more attention on artifact-rich areas of frames.
	Here, after the RF module, the spatiotemporal dependency among frames is divided into different channels.
	Thus, following the deformable convolutional operation, channel attention mechanism is used to align the spatial information along the temporal dimension.
	
	The major contributions of this paper are summarized as follows:
	\begin{itemize}
		\item We propose a recursive fusion module to exploit more spatiotemporal information in video frames in a large temporal scope but with limited additional computation cost.
		\item We develop a deformable spatiotemporal attention mechanism to guide the model on artifact-rich areas in each frame.
		\item We conduct extensive experiments on MFQE 2.0 dataset to show the superiority of the proposed method.
	\end{itemize}
	
	\begin{figure*}
		\begin{center}
			\includegraphics[width=\linewidth]{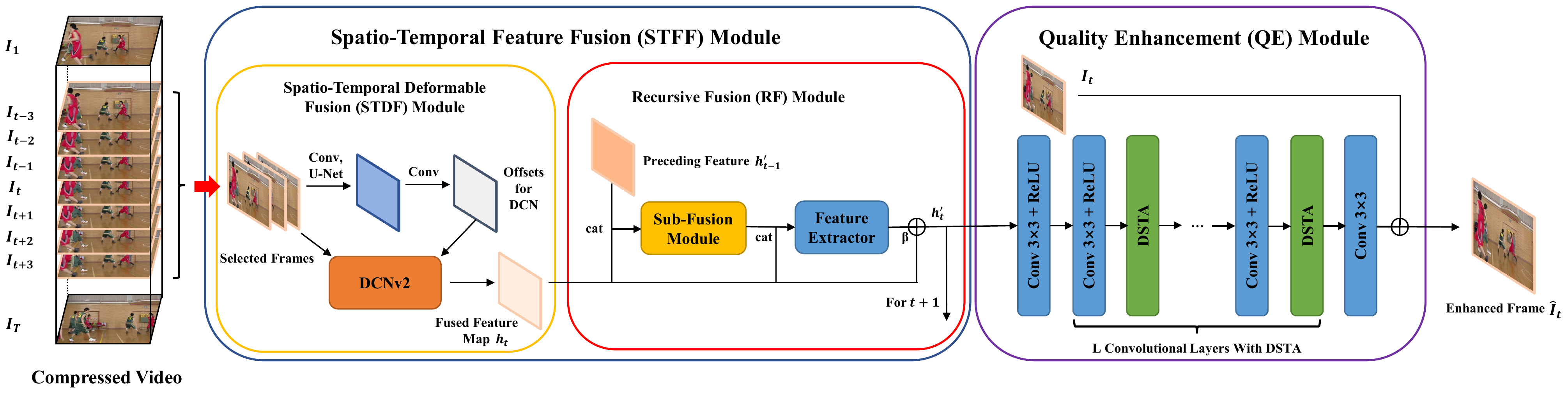}
		\end{center}
		\caption{The framework of our method, which consists of two major components: the \emph{Spatio-Temporal Feature Fusion }(STFF) module, which is designed for temporal feature fusion and the \emph{Quality Enhancement} (QE) Module, which reconstructs the fused feature. And in the STFF module, the \emph{Recursive Fusion} (RF) module is the key to fuse multiple frames in a large temporal scope.}
		\label{fig-framework}
	\end{figure*}
	
The rest of this paper is organized as follows: Section~\ref{sec-related} reviews the related work and highlights the difference between our method and the existing multi-frame based methods. Section~\ref{sec:method} presents the technical details of the proposed method. Section~\ref{sec:experiments} introduces empirical results, including performance comparison with the state of the art methods and ablation study. Section~\ref{sec:conclusion} concludes this paper.

	\section{Related Work}
	\label{sec-related}
	In this section, we review the related work of video compression artifact reduction based on deep-learning. 
	According to the domain knowledge and the number of input frames used,
	existing methods 
	can be roughly categorized into image-based approaches, single-frame based and multi-frame based approaches, respectively.
	
	\textbf{Image-based approaches.} These approaches~\cite{dong2015compression,zhang2017beyond,li2017efficient,2017Deep,yoo2018image,chen2018dpw,guo2017one,liu2018non,zhang2019residual} are proposed to solve the problem of JPEG image quality enhancement. When applied to compressed videos, these methods are fed with a single frame extracted from videos and enhance it without knowledge from video compression algorithms. For example, AR-CNN~\cite{dong2015compression} introduces four convolutional layers to reduce JPEG compression artifacts. DnCNN~\cite{zhang2017beyond} is a deep model with batch normalization and residual learning. \cite{yoo2018image,chen2018dpw} adopt wavelet/frequency domain information to enhance visual quality. NLRN~\cite{liu2018non} and RNAN~\cite{zhang2019residual} propose deeper networks with residual non-local attention mechanism to capture long-range dependencies between pixels.
	
	\textbf{Single-frame based approaches.} Among these approaches, \cite{dai2017convolutional,yang2017decoder,wang2017novel,jin2018quality,yang2018enhancing} exploit knowledge of different coding modes in video compression algorithms (\emph{e.g.} I/P/B frames), and employ special strategies to cope with them. Specifically, DS-CNN~\cite{yang2017decoder} and QE-CNN~\cite{yang2018enhancing} are proposed with two independent models to handle intra coding and inter
	coding modes, respectively. However, these methods ignore the temporal continuity of video space, which makes them hard to deal with temporal noise.
	
	\textbf{Multi-frame based approaches.} \cite{lu2018deep,lu2019deep} model the video compression artifact reduction task as a Kalman filtering procedure and capture temporal information from enhanced preceding frames to restore current frame recursively. They also incorporate quantized prediction residual in compressed code streams as strong prior knowledge. However, exploiting temporal information from only preceding frames is not enough since B-frames are compressed by using both preceding and following frames as reference. \cite{2018Multi,2019MFQE} build temporal dependency with nearby high-quality frames called PQFs. They first employ a classifier to detect PQFs, then PQFs are used as reference frames to repair non-PQFs. In implementation, they first use optical flow for motion compensation, and then design a quality enhanced network for reconstruction. Later, \cite{yang2019quality} adopts a modified convolutional LSTM to further exploit useful information of frames in a large temporal range. \cite{xu2019non,deng2020spatio} utilize non-local mechanism and deformable convolutional network to capture spatiotemporal dependency in multiple adjacent frames, respectively. More recently, \cite{xu2021boosting} employs reference frame proposals and fast Fourier transformation (FFT) loss to select high-quality reference frames and focuses on high-frequency information restoration.
	
	\textbf{Difference between our method and existing multi-frame based approaches.} Exploiting spatiotemporal information has become the mainstream of recent video enhancement works. However, existing methods are limited by the narrow temporal scope, which makes them fail to explore sufficient information from the videos.
	In our paper, a novel recursive fusion module is proposed to take advantage of both the preceding hidden state and the current feature to enlarge the receptive field and improve restoration performance. Besides, traditional non-local attention is inefficient in this task due to the huge computational cost, and treats different areas in each frame equally. Therefore, we further develop an efficient deformable spatiotemporal attention (DSTA) module to capture the artifact-rich areas and make the model pay more attention on these areas, which achieves much better restoration performance.

	\section{METHOD}\label{sec:method}
	In this section, we present the technical details of our method. 
	The architecture of our method is illustrated in Fig.~\ref{fig-framework}. Before the detailed description, we first formulate the video compression artifact reduction problem and give an overview of the method.
	
	
	\subsection{Problem Formulation and Method Overview}
	Given a compressed video $V=[I_1, I_2, \cdots, I_{T}]$ that is composed of $T$ frames, where $I_t \in \mathbb{R}^{C\times N}$ is the compressed frame at time $t$. Here, $C$ is the number of channels of a single frame, $N=H \times W$ is the collapsed spatial dimension of height $H$ and width $W$ for the sake of notation clarity. The task of video compression artifact reduction is to produce the enhanced frame $\hat{I}_t$ at time $t$ based on the input compressed video $V$.
	
	As discussed in Sec.~\ref{sec-related}, existing methods utilize different sampling strategies to take frames as input and then enhance the target frame.
	In our method, we select $2R+1$ consecutive neighboring frames as input to obtain the enhanced $\hat{I}_t$, these input frames are denoted by $X_t=\{I_{t-R},..,,I_{t-1},I_t,I_{t+1},...,I_{t+R}\}$. In other words, the input data at time $t$ consist of $R$ preceding frames, $R$ succeeding frames and the target frame $I_t$.
	Then formally, the enhanced frame $\hat{I}_t$ can be generated by
	\begin{equation}
		\label{eq-total}
		\begin{split}
			h^{'}_t &= \mathcal{F}_{\theta}(X_t,h^{'}_{t-1}),\\
			\hat{I}_t &=\mathcal{F}_{\phi}(h^{'}_t),
		\end{split}	
	\end{equation}
	where $\mathcal{F}_\theta (\cdot)$ represents the \emph{Spatio-Temporal Feature Fusion}~(STFF) module~(shown in the blue box in Fig.~\ref{fig-framework}), and $\mathcal{F}_\phi (\cdot)$ is the \emph{Quality Enhancement}~(QE) module~(shown in the purple box in Fig.~\ref{fig-framework}). $\theta$ and $\phi$ are the learnable parameters of the corresponding STFF and QE modules. $h^{'}_t \in \mathbb{R}^{F\times N}$ is the hidden state at time $t$ output by the \emph{Recursive Fusion}~(RF) module~(shown in the red box in Fig.~\ref{fig-framework}) in STFF, where $F$ indicates the number of channels.
	
	Overall, as illustrated in Fig.~\ref{fig-framework}, the STFF module is to fuse spatiotemporal information from multiple frames, where the \emph{Spatio-Temporal Deformable Fusion}~(STDF) module combines $2R+1$ frames and output the fused feature $h_t$. Following STDF, the \emph{Recursive Fusion} (RF) module produces the hidden state $h^{'}_t$ with the fused feature $h_t$ and the preceding hidden state $h^{'}_{t-1}$. Afterward, the hidden state $h^{'}_t$ goes through the QE module to generate the residual for the final enhanced frame. In our method, we employ similar architecture and settings of the STDF module in \cite{deng2020spatio}. Thus, in what follows, we only elaborate the details of the RF module in Sec.~\ref{sec-rf}, and the QE module in Sec.~\ref{sec-DA}.
	
	
	Note that there are alternatives to the proposed RF module with RNNs. Actually, we can figure out at least three ways to apply RNNs to our problem: (a) Replacing STFF by RNNs in Fig~\ref{fig-framework}; (b) Replacing STDF by RNNs; (c) Replacing RF by RNNs. For (a), it is similar to the solution in \cite{xu2019non}, which consumes too much time. As for (b), STDF~\cite{deng2020spatio} takes 2$R$+1 frames as input and fuses them in parallel, which is more efficient than one-by-one fusion via RNNs. For (c), we empirically tested ConvLSTM~\cite{xingjian2015convolutional}, and the experimental results show that ConvLSTM cannot fully exploit spatiotemporal information (See Sec~\ref{sec-abstudy}). Therefore, we discard these RNNs schemes in our architecture.

	\subsection{Recursive Fusion Module}\label{sec-rf}
	The \emph{Recursive Fusion} (RF) module is proposed to refine the fused feature $h_t$ from STDF by introducing the preceding hidden state $h^{'}_{t-1}$. As designed in \cite{deng2020spatio}, STDF utilizes an efficient U-Net based network~\cite{ronneberger2015u} to predict the offset field $\Delta \in \mathbb{R}^{(2R+1) \times 2K^2 \times N}$ for the deformable convolutional kernel,
	where $K$ is the kernel size of the deformable convolutional layer.
	Then, these position-specific offsets $\Delta$ guide the deformable convolution network to fuse the input frames efficiently.
	
	In \cite{deng2020spatio}, the fused feature $h_t$ is fed into the QE module directly, while in this paper we propose the RF module to refine $h_t$ by further exploiting spatiotemporal information of frames.
Concretely,	the core idea of the RF module is to recursively take advantage of the preceding hidden state $h^{'}_{t-1}$ as a clue to adjust the current feature $h_t$. The operation of the RF module can be formalized as
	\begin{equation}
		\label{eq-res}
		\begin{split}
			\mathcal{F}_{RF}(h_t) &= \begin{cases}
				h_t + \beta \mathcal{F}_{fe}(h_t,\mathcal{F}_{sf}(h_t, h^{'}_{t-1})) ,& t \neq 1\\
				h_1, & t = 1
			\end{cases}
		\end{split}	
	\end{equation}
	where $h_t$ is the fused feature from STDF, $\mathcal{F}_{sf}$ is the sub-fusion module which fuses the current feature $h_t$ and the preceding hidden state $h^{'}_{t-1}$. $\mathcal{F}_{fe}$ is the feature extractor which aims at learning a residual and provide additional spatiotemporal information for $h_t$. In our implementation,  $\beta$ is set as 0.2 via grid search.

	As depicted in Eq.~\ref{eq-res}, we first employ a sub-fusion module $\mathcal{F}_{sf}$ to align the feature of the preceding hidden state $h^{'}_{t-1}$ and the current feature $h_t$.
		Consider that large deformable convolutional kernels are beneficial to performance at the cost of efficiency~\cite{chan2020understanding}, the pixel-level mapping can meet the requirements of the RF module since the motion magnitude between two consecutive frames, \emph{i.e.}, $h_t$ and $h^{'}_{t-1}$, is very small.
		In implementation, we employ the deformable convolutional layer with 1$\times$1 kernel size, which means the offset field satisfies $\mathbb{R}^{F \times 2 \times N}$, where $F$ is the channel number of $h^{'}_{t-1}$.
		Then, we adopt a feature extract network to learn a residual from the aligned feature $\mathcal{F}_{sf}(h^{'}_{t-1},h_t)$ and the current feature $h_t$, and use the residual to refine $h_t$. We implement the feature extract network by stacking two convolutional layers.
		Experiments in Tab.~\ref{tab-abstudy} show that the RF module with shallow networks can achieve competitive performance. In this manner, spatiotemporal information can be fused recursively and efficiently in a large temporal scope.
		\begin{figure}
			\begin{center}
				\includegraphics[width=\linewidth]{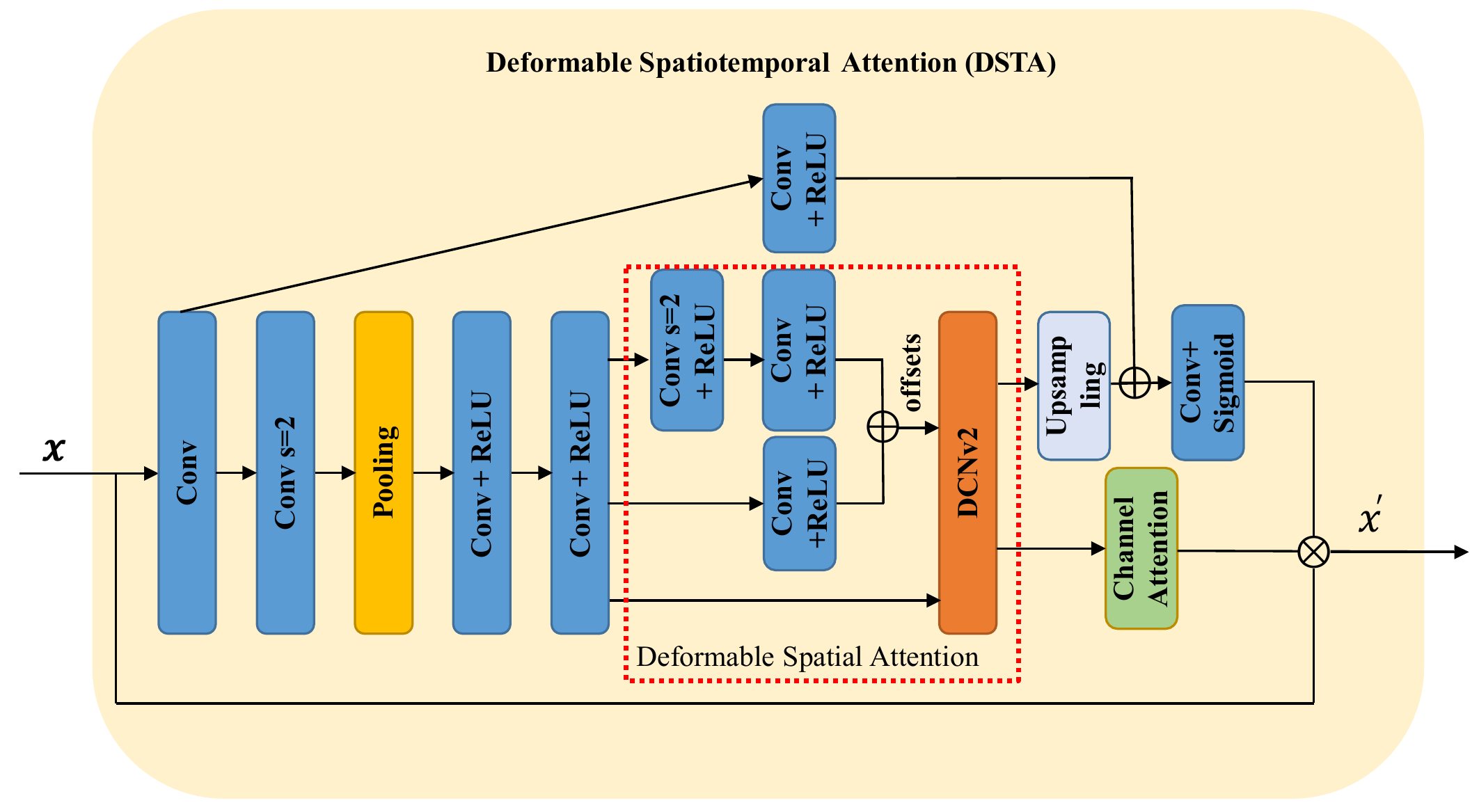}
			\end{center}
			\caption{The structure of DSTA. With no special indication, the stride of convolution kernel is 1.}
			\label{fig-dsca}
		\end{figure}
		\begin{figure}
			\begin{center}
				\includegraphics[width=\linewidth]{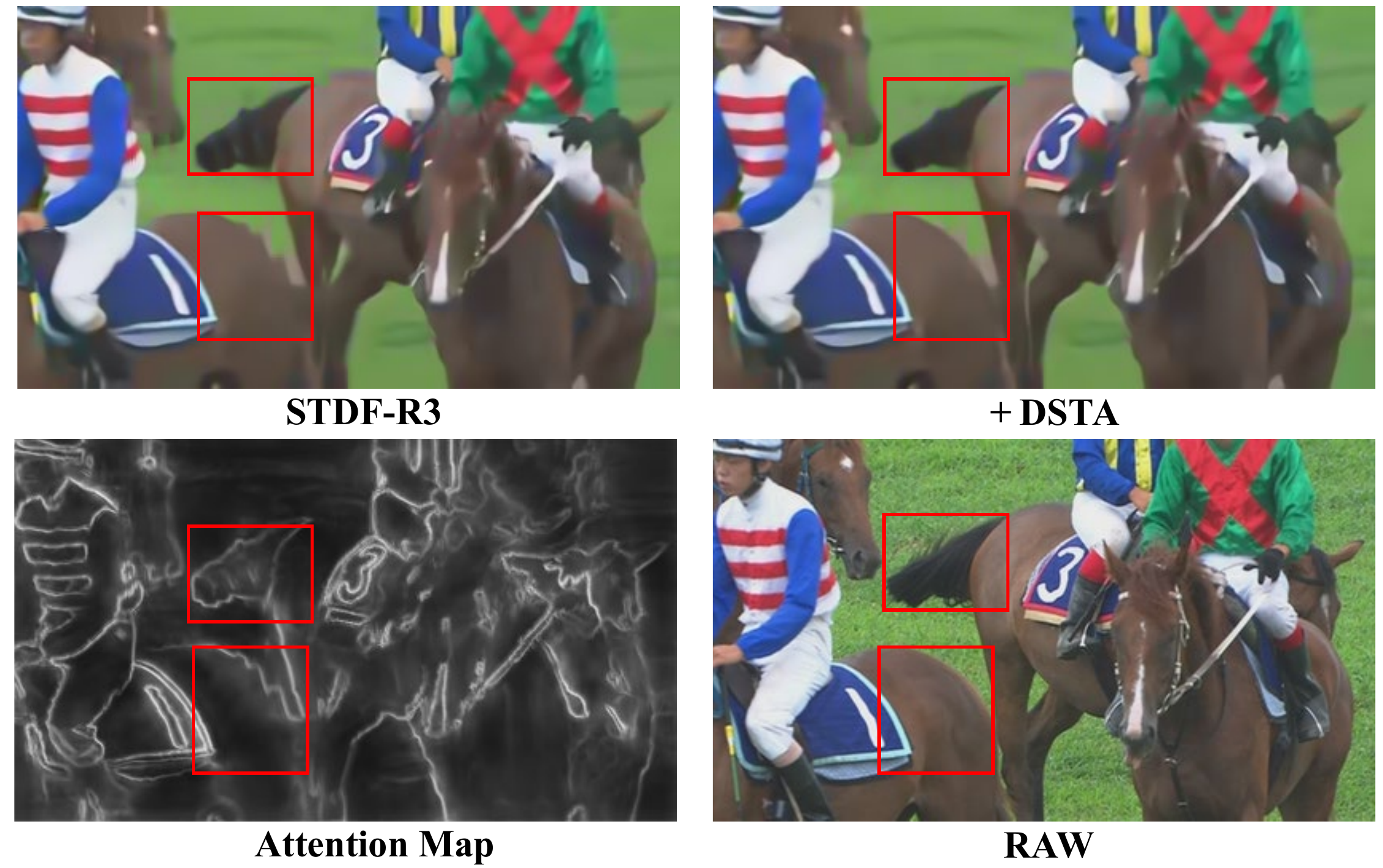}
			\end{center}
			\caption{A visual example to illustrate the effectiveness of DSTA. From top left to bottom right are the results of STDF-R3, the results of our method, the attention map in our method and the raw frame, respectively. The most distinguishable results are outlined in red boxes.}
			\label{fig-dstavis}
		\end{figure}
		\subsection{Deformable Spatiotemporal Attention}
		\label{sec-DA}
		The	QE module is essential in this task since it needs to decode the hidden state into residual, with which the high-quality frame is reconstructed. However, the QE module has received little attention in recent works.
		In this paper, the Deformable Spatiotemporal Attention (DSTA) mechanism is designed for improving the performance of the QE module. Generally, QE takes the hidden state $h^{'}_{t}$ from STFF as input and produces a residual, which is used together with the compressed frame to recover the compressed frame.
		
		The structure of the QE module can be referred to Fig.~\ref{fig-framework}, which consists of $L+2$ convolutional layers and $L$ DSTA blocks.
		Meanwhile, the structure of a DSTA block is illustrated in Fig.~\ref{fig-dsca}.
		Generally, DSTA is to make the model pay more attention to artifact-rich areas such as the boundaries of moving objects in a frame, so that the QE module can achieve better performance in these areas.
		
		In DSTA, we first adopt 1$\times$1 convolution to reduce channel number for enough receptive field and low computational cost, which is inspired by \cite{liu2020residual}.
		Then, a convolutional layer with stride length 2 is followed by a $7\times7$ pooling layer to reduce the spatial size of the feature.
		After reducing the spatial size of the feature, we utilize a multi-scale structure to predict offsets.
		Then, the deformable convolutional layer uses the predicted offsets to calculate the attention map.
		This map is then fed into the upsampling operator and channel attention module to generate the spatial attention mask and the channel weights, respectively.
		It is worth mentioning that the channel attention mechanism plays a selection role for the hidden state where different dimensions represent different temporal relationships with the target frame.
		\begin{table*}[t]
			\renewcommand\arraystretch{1.4}
			\centering
			\scriptsize
			\caption{Overall performance comparison in terms of $\Delta$PSNR (dB) and $\Delta$SSIM ($\times10^{-2}$) over the test sequences at five QPs. \\Video resolution: Class A ($2560\times1600$), Class B ($1920\times1080$), Class C ($832\times480$), Class D ($480\times240$), Class E ($1280\times720$)}
			\resizebox{0.9\textwidth}{!}{
			\begin{tabular}{|c|c|l||c@{\hspace{0.4em}} c@{\hspace{0.4em}} c@{\hspace{0.4em}} c@{\hspace{0.4em}} c@{\hspace{0.4em}} c@{\hspace{0.4em}} c @{\hspace{0.4em}} c @{\hspace{0.4em}}  c |c@{\hspace{0.4em}} c@{\hspace{0.4em}}  c |}
				
				\hline
				\multirow{2}{*}{QP} & \multicolumn{2}{c||}{\multirow{2}{*}{Approach}} & AR-CNN & DnCNN &Li \textit{et al.} &DCAD & DS-CNN & {MFQE 1.0} & {MFQE 2.0} & {STDF-R3}& {STDF-R3L}& Ours & Ours & Ours\\ [-0.3em]
				
				& \multicolumn{2}{c||}{} & {\cite{dong2015compression}}  & {\cite{zhang2017beyond}} & {\cite{li2017efficient}} & {\cite{wang2017novel}} & {\cite{yang2018enhancing}} & {\cite{2018Multi}}  & {\cite{2019MFQE}}  & \cite{deng2020spatio}& \cite{deng2020spatio} & +RF & +DSTA & RF+DSTA \\
				
				\hline
				\multirow{20}{*}{37} & \multicolumn{2}{c||}{Metrics} & \tiny PSNR / \tiny SSIM & \tiny PSNR / \tiny SSIM & \tiny PSNR / \tiny SSIM & \tiny PSNR / \tiny SSIM & \tiny PSNR / \tiny SSIM & \tiny PSNR / \tiny SSIM & \tiny PSNR / \tiny SSIM & \tiny PSNR / \tiny SSIM  & \tiny PSNR / \tiny SSIM  & \tiny PSNR / \tiny SSIM & \tiny PSNR / \tiny SSIM & \tiny PSNR / \tiny SSIM \\
				
				\cline{2-15}
				& \multirow{2}{*}{A} & \textit{Traffic}
				& 0.24 / 0.47 & 0.24 / 0.57 & 0.29 / 0.60 & 0.31 / 0.67 & 0.29 / 0.60 & 0.50 / 0.90 & 0.59 / 1.02 & 0.65 / 1.04 &0.73 / 1.15& 0.76 / 1.24 & 0.78 / 1.24 &  \textbf{0.80} / \textbf{1.28} \\
				
				& & \textit{PeopleOnStreet}
				& 0.35 / 0.75 & 0.41 / 0.82 & 0.48 / 0.92 & 0.50 / 0.95 & 0.42 / 0.85 & 0.80 / 1.37 & {0.92} / {1.57} & 1.18 / 1.82 &1.25 / 1.96& 1.34 / 2.10 & 1.39 / 2.16 & \textbf{1.44} / \textbf{2.22}\\
				
				\cline{2-15}
				& \multirow{5}{*}{B} & \textit{Kimono}
				& 0.22 / 0.65 & 0.24 / 0.75 & 0.28 / 0.78 & 0.28 / 0.78 & 0.25 / 0.75 & 0.50 / 1.13 & {0.55} / {1.18} & 0.77 / 1.47 &0.85 / 1.61& 1.02 / 1.86 & 1.01 / 1.85 &  \textbf{1.02} / \textbf{1.86}\\
				
				& & \textit{ParkScene}
				& 0.14 / 0.38 & 0.14 / 0.50 & 0.15 / 0.48 & 0.16 / 0.50 & 0.15 / 0.50 & 0.39 / 1.03 & {0.46} / {1.23} & 0.54 / 1.32&0.59 / 1.47 & 0.63 / \textbf{1.58} & 0.63 / 1.52 & \textbf{0.64} / \textbf{1.58}\\
				
				& & \textit{Cactus}
				& 0.19 / 0.38 & 0.20 / 0.48 & 0.23 / 0.58 & 0.26 / 0.58 & 0.24 / 0.58 & 0.44 / 0.88 & {0.50} / {1.00} & 0.70 / 1.23 &0.77 / 1.38& 0.81 / \textbf{1.49} & 0.81 / 1.46 & \textbf{0.83} / \textbf{1.49}\\
				
				& & \textit{BQTerrace}
				& 0.20 / 0.28 & 0.20 / 0.38 & 0.25 / 0.48 & 0.28 / 0.50 & 0.26 / 0.48 & 0.27 / 0.48 & {0.40} / {0.67} & 0.58 / 0.93&0.63 / 1.06 & \textbf{0.66} / \textbf{1.07} & 0.63 / 1.01 & 0.65 / 1.06\\
				
				& & \textit{BasketballDrive}
				& 0.23 / 0.55 & 0.25 / 0.58 & 0.30 / 0.68 & 0.31 / 0.68 & 0.28 / 0.65 & 0.41 / 0.80 & {0.47} / {0.83} & 0.66 / 1.07&0.75 / 1.23 & \textbf{0.89} / \textbf{1.42} & 0.84 / 1.37 &  0.87 / 1.40\\
				
				\cline{2-15}
				& \multirow{4}{*}{C} & \textit{RaceHorses}
				& 0.22 / 0.43 & 0.25 / 0.65 & 0.28 / 0.65 & 0.28 / 0.65 & 0.27 / 0.63 & 0.34 / 0.55 & {0.39} / {0.80} & 0.48 / 1.09 &0.55 / 1.35& \textbf{0.55} / \textbf{1.32} & 0.51 / 1.22 & 0.48 / 1.23\\
				
				& & \textit{BQMall}
				& 0.28 / 0.68 & 0.28 / 0.68 & 0.33 / 0.88 & 0.34 / 0.88 & 0.33 / 0.80 & 0.51 / 1.03 & {0.62} / {1.20} & 0.90 / 1.61 &0.99 / 1.80& 1.07 / 1.94 &1.06 / 1.91 & \textbf{1.09} / \textbf{1.97}\\
				
				& & \textit{PartyScene}
				& 0.11 / 0.38 & 0.13 / 0.48 & 0.13 / 0.45 & 0.16 / 0.48 & 0.17 / 0.58 & 0.22 / 0.73 &{0.36} / {1.18} & 0.60 / 1.60 &0.68 / 1.94& \textbf{0.71} / \textbf{2.06} &0.63 / 1.78 & 0.66 / 1.88 \\
				
				& & \textit{BasketballDrill}
				& 0.25 / 0.58 & 0.33 / 0.68 & 0.38 / 0.88 & 0.39 / 0.78 & 0.35 / 0.68 & 0.48 / 0.90 &{0.58} / {1.20} & 0.70 / 1.26&0.79 / 1.49 & \textbf{0.90} / \textbf{1.67} &0.87 / 1.66 & 0.88 / \textbf{1.67}\\
				
				\cline{2-15}
				& \multirow{4}{*}{D} & \textit{RaceHorses}
				& 0.27 / 0.55 & 0.31 / 0.73 & 0.33 / 0.83 & 0.34 / 0.83 & 0.32 / 0.75 & 0.51 / 1.13 & {0.59} / {1.43} & 0.73 / 1.75 &0.83 / 2.08& \textbf{0.88} / \textbf{2.18} & 0.85 / 2.08 & 0.85 / 2.11 \\
				
				& & \textit{BQSquare}
				& 0.08 / 0.08 & 0.13 / 0.18 & 0.09 / 0.25 & 0.20 / 0.38 & 0.20 / 0.38 & -0.01 / 0.15 & {0.34} / {0.65} & 0.91 / 1.13&0.94 / 1.25 & 0.95 / 1.34 & 0.96 / 1.23 & \textbf{1.05} / \textbf{1.39}\\
				
				& & \textit{BlowingBubbles}
				& 0.16 / 0.35 & 0.18 / 0.58 & 0.21 / 0.68 & 0.22 / 0.65 & 0.23 / 0.68 & 0.39 / 1.20 & {0.53} / {1.70} & 0.68 / 1.96 &0.74 / 2.26& 0.77 / \textbf{2.41} & 0.76 / 2.28 & \textbf{0.78} / 2.40 \\
				
				& & \textit{BasketballPass}
				& 0.26 / 0.58 & 0.31 / 0.75 & 0.34 / 0.85 & 0.35 / 0.85 & 0.34 / 0.78 & 0.63 / 1.38 & {0.73} / {1.55} & 0.95 / 1.82 &1.08 / 2.12& \textbf{1.15} / \textbf{2.24} & 1.09 / 2.18 & 1.12 / 2.23 \\
				
				\cline{2-15}
				& \multirow{3}{*}{E} & \textit{FourPeople}
				& 0.37 / 0.50 & 0.39 / 0.60 & 0.45 / 0.70 & 0.51 / 0.78 & 0.46 / 0.70 & 0.66 / 0.85 & {0.73} / {0.95} & 0.92 / 1.07  &0.94 / 1.17& 1.02 / 1.28 & 1.11 / 1.35 & \textbf{1.13} / \textbf{1.36} \\
				
				& & \textit{Johnny}
				& 0.25 / 0.10 & 0.32 / 0.40 & 0.40 / 0.60 & 0.41 / 0.50 & 0.38 / 0.40 & 0.55 / 0.55 & {0.60} / {0.68} & 0.69 / 0.73 &0.81 / 0.88& 0.86 / 0.91 & 0.88 / 0.90 & \textbf{0.90} / \textbf{0.94}\\
				
				& & \textit{KristenAndSara}
				& 0.41 / 0.50 & 0.42 / 0.60 & 0.49 / 0.68 & 0.52 / 0.70 & 0.48 / 0.60 & 0.66 / 0.75 & {0.75} / {0.85} & 0.94 / 0.89 &0.97 / 0.96& 1.08 / 1.11 & 1.16 / 1.14 & \textbf{1.19} / \textbf{1.15} \\
				
				\cline{2-15}
				& \multicolumn{2}{c||}{Average}
				& 0.23 / 0.45 & 0.26 / 0.58 & 0.30 / 0.66 & 0.32 / 0.67 & 0.30 / 0.63 & 0.46 / 0.88 & {0.56} / {1.09} & 0.75 / 1.32 & 0.83 / 1.51 & 0.89 / \textbf{1.62} & 0.89 /1.57 & \textbf{0.91} / \textbf{1.62} \\
				
				\hline
				\hline
				42 & \multicolumn{2}{c||}{Average}
				& 0.29 / 0.96 & 0.22 / 0.77 & 0.32 / 1.05 & 0.32 / 1.09 & 0.31 / 1.01 & 0.44 / 1.30 & {0.59} / {1.65} & -- / --  &-- / --& 0.73 / 1.97 & 0.77 / 2.12 & \textbf{0.82} / \textbf{2.20} \\
				
				\hline
				32 & \multicolumn{2}{c||}{Average}
				& 0.18 / 0.19 & 0.26 / 0.35 & 0.28 / 0.37 & 0.32 / 0.44 & 0.27 / 0.38 & 0.43 / 0.58 & {0.516} / {0.68} & 0.73 / 0.87  & 0.86 / 1.04& 0.85 / 1.05 & 0.83 / 1.05 & \textbf{0.87} / \textbf{1.07} \\
				
				\hline
				27 & \multicolumn{2}{c||}{Average}
				& 0.18 / 0.14 & 0.27 / 0.24 & 0.30 / 0.28 & 0.32 / 0.30 & 0.27 / 0.23 & 0.40 / 0.34 & {0.49} / {0.42} & 0.67 / 0.53   & 0.72 / 0.57& 0.80 / 0.67 & 0.80 / 0.67 & \textbf{0.82} / \textbf{0.68} \\
				
				\hline
				22 & \multicolumn{2}{c||}{Average}
				& 0.14 / 0.08 & 0.29 / 0.18 & 0.30 / 0.19 & 0.31 / 0.19 & 0.25 / 0.15 & 0.31 / 0.19 & {0.46} / {0.27} & 0.57 / 0.30 &0.63 / 0.34 & 0.73 / 0.40 & 0.74 / 0.41 & \textbf{0.76} / \textbf{0.42} \\
				\hline
				
			\end{tabular}
			
			}
			\label{tab-sota-results}
		\end{table*}
	 Thus, the channel attention mechanism can discriminate the useful temporal information through weighted sum over the channels.
		Finally, the input of the DSTA block is multiplied with the spatial map and the channel weights.
		
		Note that non-local attention mechanism~(see \cite{zhang2019residual}) may bring more performance improvement, but its huge amount of computation increases time cost for restoration, which is unacceptable and not in line with our motivation. Our DSTA module achieves a good balance between computation and performance, and takes both temporal and spatial dependencies into consideration to get better performance. As shown in Fig.~\ref{fig-dstavis}, the proposed attention block pays more attention to moving areas, therefore guides the QE module to put more effort to restore them.

		\subsection{Training Scheme}
		\label{sec-twostage}
		To further boost the performance of our method, we train the model in two stages. In the first stage, we remove the RF module and focus on STDF and DSTA modules. When these two modules converge, we reduce their learning rates and add the RF module into training. In the second phase, we utilize video clips as input to train our RF module. This training strategy can reduce the training time. In contrast, the convergence speed of one-stage training strategy is quite slower than ours.
		In both stages, we employ the Charbonnier Loss \cite{charbonnier1994two} to optimize the model:
		\begin{equation}
			\mathcal{L}= \sqrt{(\hat{I}_t - \overline{I}_t)^2+\epsilon}
		\end{equation}
		where $\hat{I}_t$ is the enhanced frame, $\overline{I}_t$ is the ground truth, and $\epsilon$ is set to $10^{-6}$ in our paper.
		
		\section{Experiments}\label{sec:experiments}
		In this section, we conduct extensive experiments to evaluate the effectiveness and superiority of our proposed approach. Our evaluation consists of two parts: (1) comparison with state-of-the-art methods, and (2) Ablation study on the effects of different sub-modules, all on the MFQE 2.0 dataset with five QPs.
		
		\subsection{Datasets}
		Following \cite{2019MFQE,deng2020spatio, xu2021boosting}, we conduct our experiments on MFQE 2.0 dataset. It consists of 126 video sequences collected form Xiph.org \cite{Xiph}, VQEG \cite{VQEG} and JCT-VC \cite{bossen2011common}. The video sequences contained in MFQE 2.0 are at large
		range of resolutions: SIF (352$\times$240), CIF (352$\times$288), NTSC (720$\times$486), 4CIF (704$\times$576), 240p (416$\times$240), 360p (640$\times$360), 480p (832$\times$480), 720p (1280$\times$720), 1080p (1920-$\times$1080), and WQXGA (2560$\times$1600). For a fair comparison, we follow the settings in \cite{2019MFQE,deng2020spatio, xu2021boosting}: 108 of them are taken for training and the remaining 18 for test. All sequences are encoded in HEVC Low-Delay-P (LDP) configuration, using HM 16.20 with \emph{QP}=22, 27, 32, 37 and 42~\cite{2019MFQE}.
		
		\subsection{Implementation Details}
		In this paper, we take the state-of-the-art method STDF-R3 in \cite{deng2020spatio} as our baseline, which means $R=3$ for STDF in our paper. In the training phase, we randomly crop 112$\times$112 clips from the raw videos and the corresponding compressed videos as training samples with setting batch size 32. We also adopt flip and rotation as data augmentation strategies to further expand the dataset. In the first training stage, the model is trained by the Adam~\cite{kingma2014adam} optimizer with an initial learning rate of $10^{-4}$, which is decreased by half when $60\%$ and $90\%$ iterations are reached.
		In the second training stage, we set the learning rate of STDF and QE to $10^{-5}$ and split each video in the training set into several video clips. Each video clip contains 15 frames.
		It is worth mentioning that in the first training stage, we calculate the loss frame by frame, while in the second training stage, we calculate the loss over the entire clip to better exploit the characteristic of the RF module. Our model is trained on 4 NVIDIA GeForce RTX 3090 GPUs with PyTorch1.8.
		
		For evaluation, following the settings of \cite{deng2020spatio, xu2021boosting}, we only report quality enhancement on Y-channel in YUV/YCbCr space. We adopt improvement over compressed Peak Signal-to-Noise Ratio ($\Delta$PSNR) and Structural Similarity  ($\Delta$SSIM)~\cite{wang2004image} to evaluate the quality enhancement performance. Taking $\Delta$PSNR for example, we use the difference between the enhanced PSNR~(calculated from the enhanced video $\hat{V}$ and the ground truth $\overline{V}$) and the compressed PSNR~(calculated from the compressed video $V$ and the ground truth $\overline{V}$) to measure the performance of the method, i.e., $\Delta PSNR=PSNR(\hat{V},\overline{V})-PSNR(V,\overline{V})$.

		\subsection{Comparison with State of the art Methods}
		To demonstrate the advantage of our method, we compare our method with the state-of-the-art approaches, including image-based \cite{dong2015compression,zhang2017beyond,li2017efficient}, singe-frame based \cite{wang2017novel,yang2018enhancing} and multi-frame approaches \cite{2018Multi,2019MFQE,deng2020spatio}. To fully validate the effectiveness of the modules in our method, we also apply RF and DSTA to STDF-R3, respectively. Results of existing methods are cited from \cite{2019MFQE,deng2020spatio}.
		
		\textbf{Overall performance.} Results of PSNR / SSIM improvement are presented in Tab.~\ref{tab-sota-results}. In order to verify the effectiveness of different components, we also report the performance when using only RF and DSTA, respectively.
		From Tab.~\ref{tab-sota-results} we can see that multi-frame based methods perform better than all image-based approaches due to the benefit of utilizing spatiotemporal information. While our method performs better than all existing methods at five QPs, which verifies the effectiveness and superiority of the proposed method. Moreover, the STDF-R3 with only RF or DSTA also outperforms the state-of-the-art approaches at four QPs.
		
		\begin{table}
			\centering
			\caption{Inference speed and parameter size comparison between our method and some mainstream methods. For a fair comparison, all methods are retested on a NVIDIA RTX 3090. The results are reported by frame per second (FPS) and \#Params (K).}
			\centering
			\begin{tabular}{c|c|c|c|c|c}
				\toprule
				Method  & 480p & 720p & 1080p & WQXGA & \#Params(K)\cr
				\midrule
				STDF-R3  & 25.1 & 11.2 & 5.0 &2.5 & 365 \\
				STDF-R3L  & 10.3 & 4.5 & 2.0 &1.0& 1275 \\
				+RF  & 12.6 & 5.6 & 2.5 &1.2& 770 \\
				+DSTA  & 17.8 & 8.1 & 3.7 &1.8& 840 \\
				RF+DSTA & 10.5 & 4.7 & 2.1 &1.1& 1250 \\
				\bottomrule
			\end{tabular}
			\label{tab-speed}
		\end{table}
		
		\textbf{Speed and parameter size comparison.} We also compare inference speed and parameter size between our method and some existing methods. As shown in Tab.~\ref{tab-speed}, our method outperforms the state-of-the-art method STDF-R3L in terms of both inference speed and parameter size. Specifically, our method gets an averaged $\Delta$PSNR improvement of 8.8\% (from 0.83 to 0.91, see Tab.~\ref{tab-sota-results}) and boosts the inference FPS by 0.1$\sim$0.2 (improved by 2$\sim$10\%, see Tab.\ref{tab-speed}), while our model has less parameters than STDF-R3L (1250k vs. 1275k, see Tab.~\ref{tab-speed}), which demonstrates the effectiveness and superiority of our proposed method. Furthermore,
		by checking both Tab.~\ref{tab-sota-results} and Tab.~\ref{tab-speed}, we can see that even using only RF or DSTA, we can still achieve better performance at four QPs than STDF-R3L in terms of restoration effect, speed and parameter size.
		
		\begin{table}
			\caption{Averaged PVD/SD of test sequences for PSNR at \emph{QP}=27, 32, 37 and 42.}
			\centering
			
			\begin{tabular}{c|c@{\hspace{0.6em}} c@{\hspace{0.6em}} c@{\hspace{0.6em}} c}
				\toprule
				Method & QP27 & QP32 & QP37 & QP42 \cr
				\midrule
				HEVC & 1.07 / 0.83 & 1.38 / 0.82 & 1.42 / 0.79 & 1.21 / 0.74 \\
				AR-CNN & 1.07 / 0.83 & 1.38 / 0.82 & 1.44 / 0.80 & 1.24 / 0.75 \\
				DnCNN & 1.06 / 0.83 & 1.40 / 0.83 & 1.44 / 0.80 & 1.24 / 0.75 \\
				Li \textit{et al.}\cite{li2017efficient} & 1.06 / 0.83 & 1.38 / 0.83 & 1.44 / 0.80 & 1.24 / 0.76 \\
				DCAD & 1.07 / 0.83 & 1.39 / 0.83 & 1.45 / 0.80 & 1.26 / 0.76 \\
				DS-CNN & 1.07 / 0.83 & 1.39 / 0.83 & 1.46  / 0.80 & 1.24 / 0.75 \\
				MFQE 1.0 & 0.84 / 0.81 & 1.07 / 0.77 & 1.05 / 0.73 & 0.82 / 0.69 \\
				MFQE 2.0 & 0.77 / 0.74 & 0.98 / 0.70 & 0.96 / 0.67 & 0.74 / 0.62 \\
				Ours & \textbf{0.59} / \textbf{0.45} & \textbf{0.77} / \textbf{0.41} & \textbf{0.71} / \textbf{0.39} & \textbf{0.64} / \textbf{0.36}\\
				\bottomrule
			\end{tabular}
			
			\label{tab-SDPVD}
		\end{table}
		\begin{figure}
			\centering
			\includegraphics[width=\linewidth]{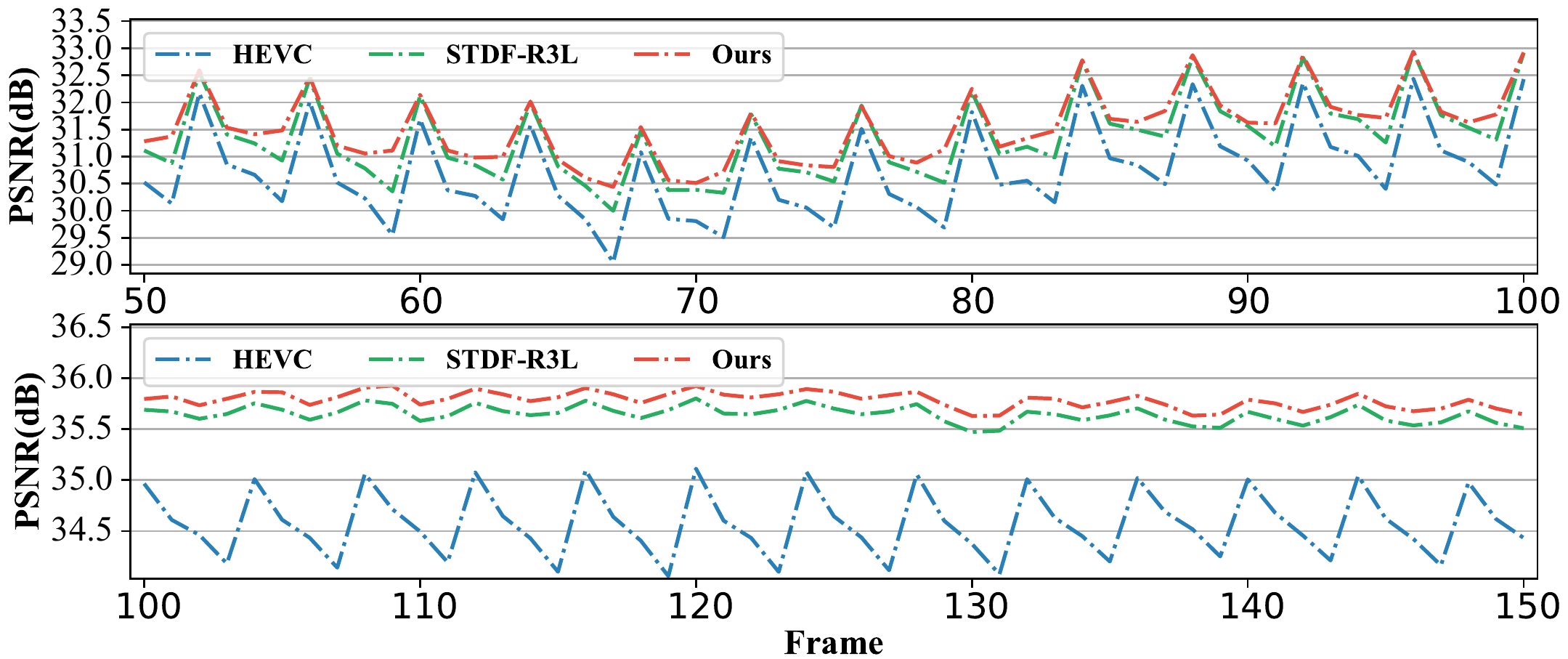}
			\caption{PSNR curves of HEVC baseline, STDF-R3L and ours on two test sequences at \emph{QP}=37. Video index of top: \emph{BasketballDrive} and bottom: \emph{FourPeople}.}
			\label{fig-curve}
		\end{figure}

		\textbf{Quality fluctuation.}
		Quality fluctuation is another observable measurement for the overall quality of enhanced videos. Drastic quality fluctuation of frames accounts for severe texture shaking and degradation of the quality of experience. We evaluate the fluctuation by Standard Deviation~(SD) and Peak-Valley Difference~(PVD) of each test sequence as in \cite{yang2018enhancing,xu2019non, 2019MFQE}. The averaged PVD and SD over all test sequences for our method and major existing methods are presented in Tab.~\ref{tab-SDPVD}. We can see that our method has the smallest averaged PVD and SD, which shows that our method is more stable than the other approaches.
		
		Fig.~\ref{fig-curve} shows two groups of PSNR curves, each group consists of three PSNR curves of a test sequence compressed by HEVC and the two corresponding sequences enhanced by STDF-R3L and our method. Comparing with STDF-R3L, our method has larger PSNR improvement over the compressed frames, but smaller fluctuation.
		\begin{figure}
			\centering
			\includegraphics[width=0.9\linewidth]{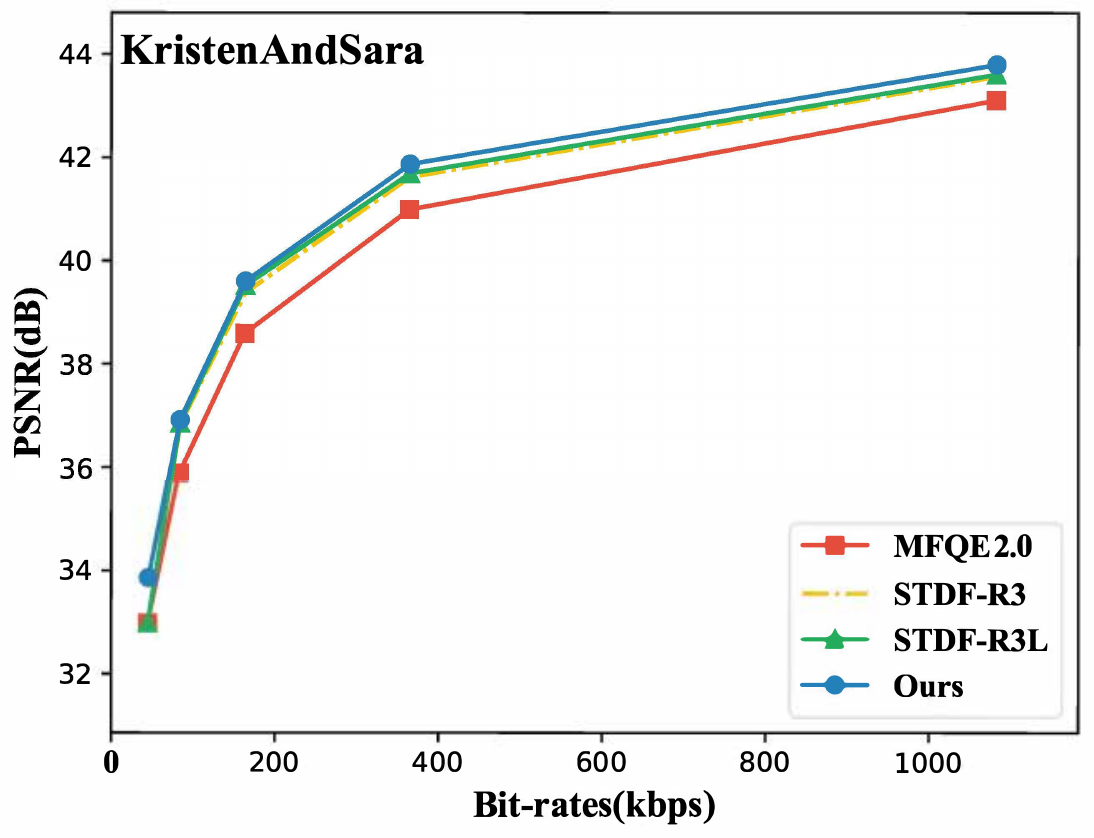}
			\caption{Rate-distortion performance on video index: \textit{KristenAndSara}. The proposed method (blue curve) outperforms the other approaches at various bit-rates. }
			\label{fig-rd}
		\end{figure}
		
		\textbf{Rate-distortion.}
		Here, we evaluate the rate-distortion of our method and compare it with state-of-the-art approaches. For simplicity of illustration, we present only the results of compressed videos, the enhanced results of two state-of-the-art methods~(MFQE 2.0 and STDF-R3L) and our method in Fig.~\ref{fig-rd}. Due to the lack of data in \cite{deng2020spatio}, we do not show the results of STDF-R3L at $QP$=42.
		From Fig.~\ref{fig-rd} we can see that for similar bit-rate, our method gets larger PSNR than the other approaches, which indicates that our method performs better than the state-of-the-art approaches in terms of rate-distortion.
		

		\begin{table}
			\caption{Ablation study at \emph{QP}=37 with STDF-R3 as baseline. Results of $\Delta$PSNR (dB) and $\Delta$SSIM ($\times 10^{-2}$) are presented.}
			\begin{tabular}{ccc|c}
				\toprule
				RF & DSTA  & Two-Stage & $\Delta$PSNR / $\Delta$SSIM \cr
				\midrule
				- & - & - & 0.78 / 1.46   \\ 
				- & $\checkmark$ & - & 0.89 / 1.57   \\ 
				$\checkmark$ & - & - & 0.75 / 1.42  \\ 
				$\checkmark$ & - & $\checkmark$ & 0.89 / 1.62  \\ 
				$\checkmark$ & $\checkmark$ & $\checkmark$ & 0.91 / 1.62  \\ 
				\bottomrule
			\end{tabular}
			\label{tab-abstudy}
		\end{table}
		
		\subsection{Ablation Study}
		\label{sec-abstudy}
		To validate the effect of our modules, we take STDF-R3 as the baseline and insert different module combinations into the baseline. For a fair comparison, we retrain STDF-R3 and adopt similar experimental settings. Tab.~\ref{tab-abstudy} presents the results of STDF-R3 and STDF-R3 with different module combinations. For simplicity, we abbreviate two-stage training strategy as Two-Stage.
		
		\begin{figure}
			\begin{center}
				\includegraphics[width=\linewidth]{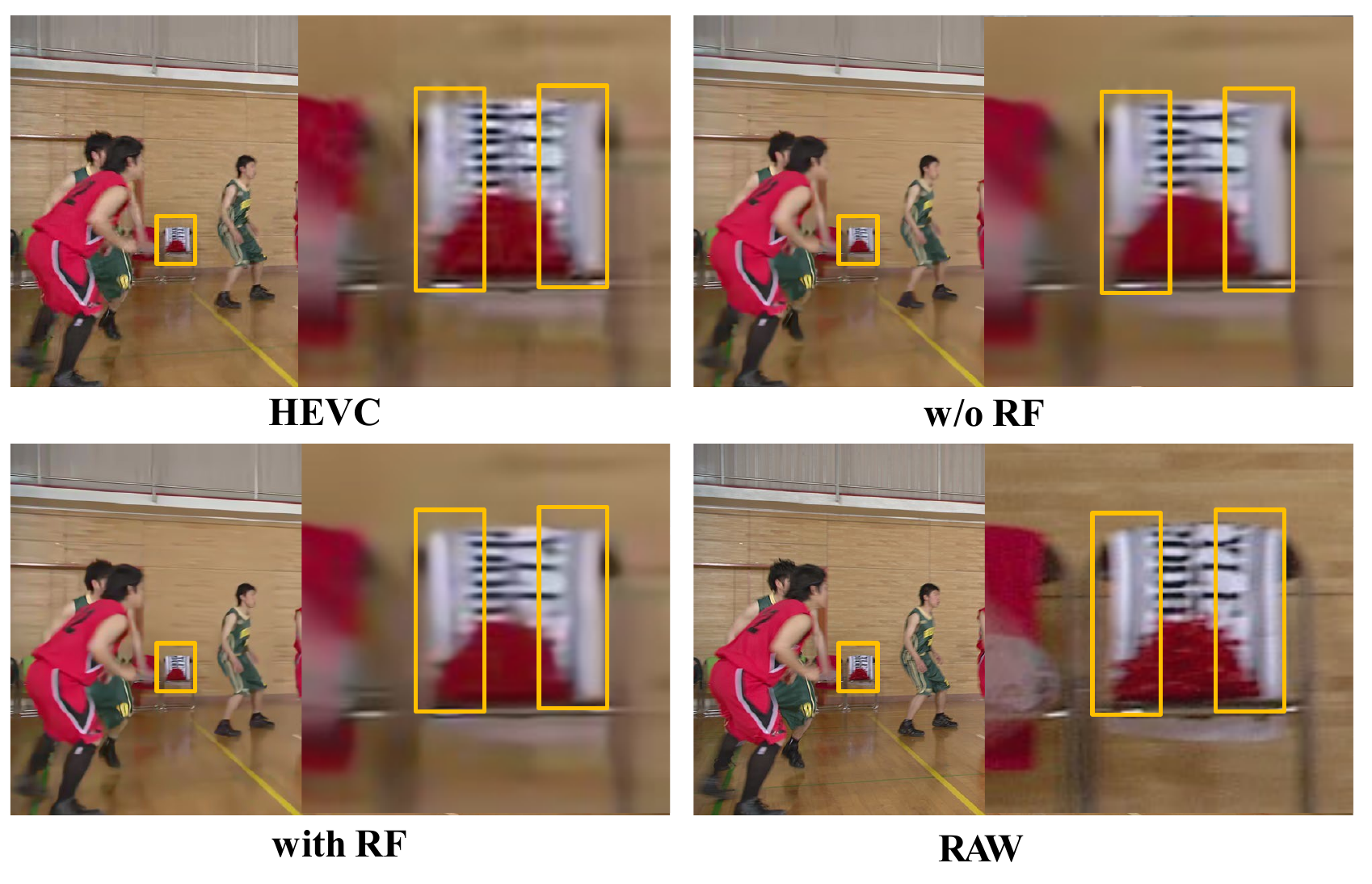}
			\end{center}
			\caption{A visual example illustrating the effectiveness of RF. With the help of RF, we can catch the spatiotemporal dependency from preceding frames in a large temporal scope, which leads to the success recovery of the bag (in the orange box).}
			\label{fig-abvis}
		\end{figure}
		\begin{figure}
			\begin{center}
				\includegraphics[width=\linewidth]{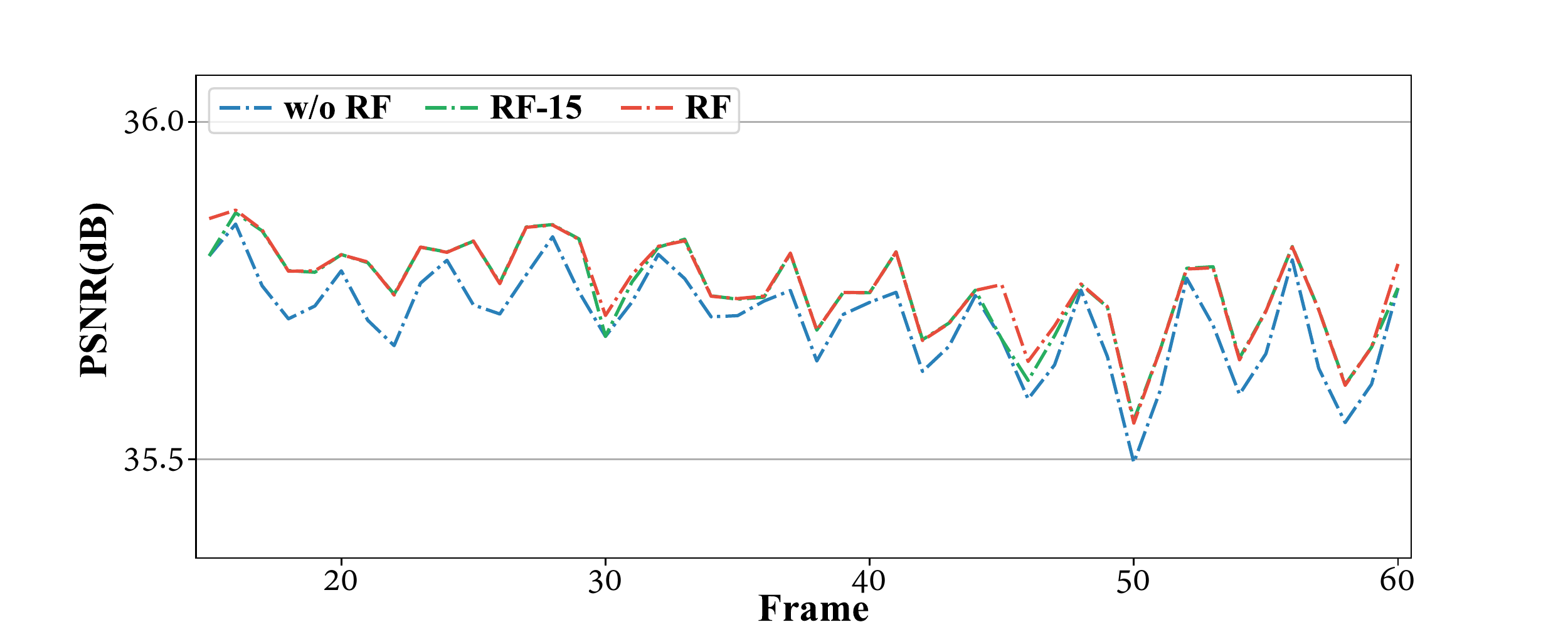}
			\end{center}
			\caption{PSNR curves of methods with different RF settings: w/o RF, RF-15 and RF on video index \emph{FourPeople}.}
			\label{fig-rf-inference}
		\end{figure}
		\textbf{Effect of Recursive Fusion.}
		Here, we evaluate the effectiveness of our RF module. As shown in the 4th raw in Tab.~\ref{tab-abstudy}, the proposed RF module with two-stage training strategy obtains a $\Delta$PSNR of 0.89, 0.11 higher than the baseline method. Visual examples in Fig.~\ref{fig-abvis} also show that the RF module brings benefit by learning to recover missed details from the preceding frames. In order to better illustrate the effectiveness of the RF module, we also train an STDF-R4 model, which takes 2 extra reference frames to conduct enhancement. STDF-R4 gets a $\Delta$PSNR of 0.79, while our RF module achieves a $\Delta$PSNR of 0.89, which again proves the effectiveness of the RF module.
		\begin{figure*}
			\centering
			\includegraphics[width=0.88\linewidth]{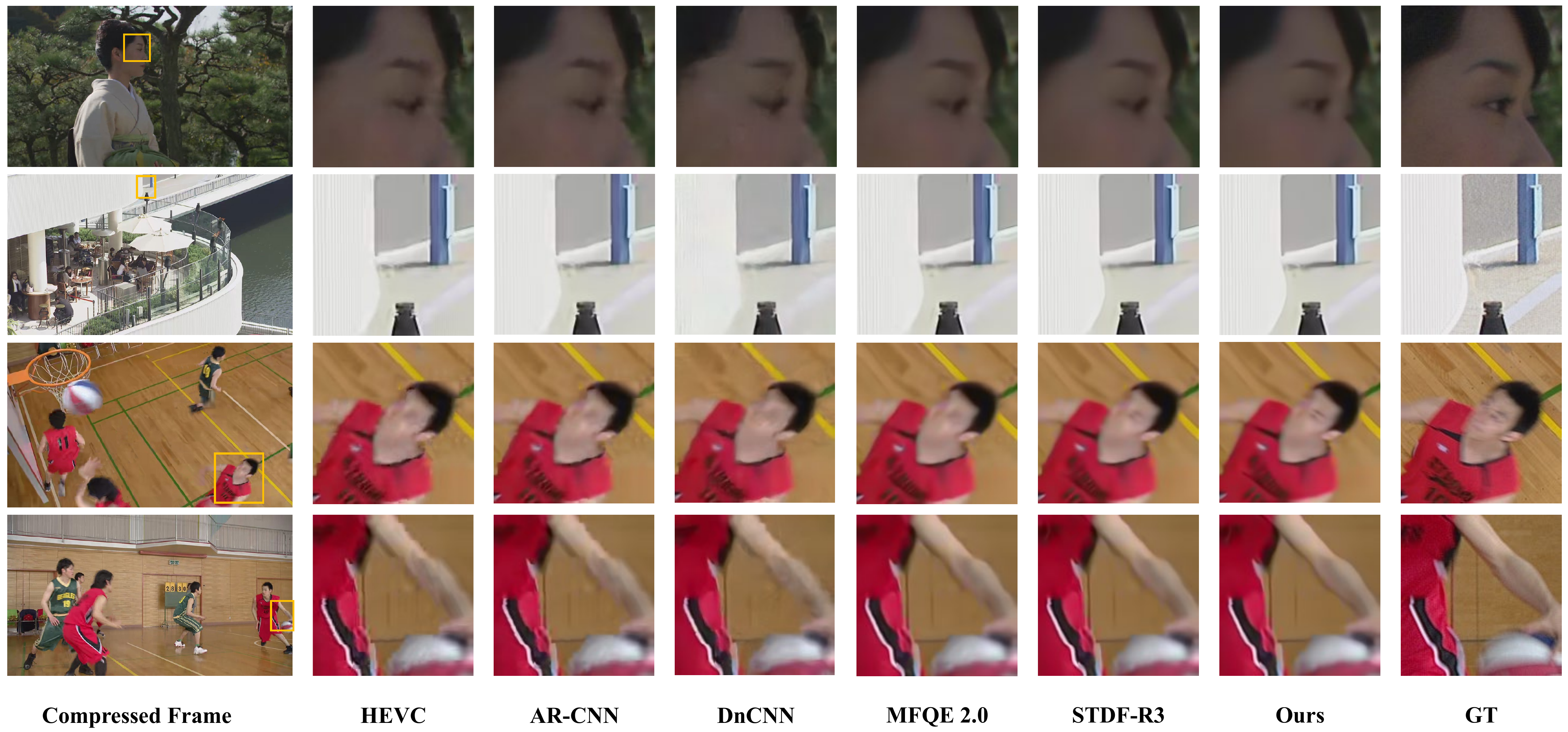}
			\caption{Examples for qualitative comparison at $QP$=37. Video indexes (from top to bottom) are: \emph{Kimono}, \emph{BQTerrace}, \emph{BasketballDrive}, \emph{BasketballDrill}. Our method gets clearer eyebrows of the woman, shadow of the pillar, eyebrow of the player, and left arm of the player, respectively.}
			\label{fig-case-qp37}
		\end{figure*}
		We notice that there exists inconsistency between our training and inference phases: We train our model with video clips of length 15, and during inference, we use RF to generate the entire video recursively. To better verify whether our model has the ability to exploit the spatiotemporal dependency in long temporal scope, we design three variants: without RF, with RF, and with RF-15, where RF-15 means that in both training and inference, we split each video into multiple 15-frame video clips and enhance them one by one. Fig.~\ref{fig-rf-inference} shows the PSNR curves of various RF variants. Both RF and RF-15 outperform w/o RF. However, RF-15 performs worse at 
		the starting frames in each clip than RF, which further verifies the effectiveness of introducing far away preceding frame information.
		
		We also replace RF with ConvLSTM~\cite{xingjian2015convolutional} and compare it with RF under the same experimental setting (Two-Stage\&QP=37). We get a 0.78 $\Delta$PSNR, much lower than 0.89 of RF, which also proves the superiority of RF.

		\textbf{Effect of Deformable Spatiotemporal Attention.} From the results in Tab.~\ref{tab-abstudy}, we can conclude that DSTA contributes a lot to performance improvement.
		For a deep investigation, we consider an additional ablation study about the branch configuration of the DSTA module.
		Intuitively, there are three variants of the proposed DSTA module: DSTA without DA (removing the deformable attention part and using basic convolutional layers to generate attention map), DSTA without CA (removing the channel attention part), and DSTA (using both branches).
		Experimental results are presented in Tab.~\ref{tab-dsca}. 
		We can see that both the single-branch attention settings are able to boost the performance of STDF-R3, but all perform worse than DSTA.
		
		\begin{table}
			\caption{Ablation study on DSTA configuration at \emph{QP}=$37$. We also report the results of parameter size and inference speed (FPS) for processing 1080p videos.}
			\centering
			\begin{tabular}{c|c|c|c}
				\toprule
				Method & $\Delta$PSNR / $\Delta$SSIM & \#Params(K) & FPS \cr
				\midrule
				STDF-R3 & 0.78 / 1.46  & 360 & 4.96 \\
				DSTA w/o DA &  0.85 / 1.55  & 400 & 3.86\\ 
				DSTA w/o CA  &  0.88 / 1.63  & 831 & 3.71\\ 
				DSTA  &  0.89 / 1.57  & 840 & 3.65\\ 
				\bottomrule
			\end{tabular}
			\label{tab-dsca}
		\end{table}
		
		\textbf{Effect of Two-Stage Training.}
		Here, we check the necessity of two-stage training strategy. For a fair comparison, we use the same optimizer and learning rate decay settings to train model with / without two-stage strategy. 
		Results of models trained with / without Two-Stage are presented in Tab.~\ref{tab-abstudy}.
		We can see that Two-Stage gets a 0.14 / 0.2$\times 10^{-2}$ $\Delta$PSNR/$\Delta$SSIM improvement~(the 4th row vs. the 3rd row), which shows the effectiveness of the two-Stage strategy. 

		\subsection{Qualitative Comparison}
		We also conduct qualitative comparison and present several visual examples at $QP$=37 in Fig.~\ref{fig-case-qp37}. Here, we compare our method with four major existing approaches: AR-CNN, DnCNN, MFQE 2.0 and baseline STDF-R3.
		As shown in Fig.~\ref{fig-case-qp37},
		the compressed patches (in the 2nd column) suffer from severe compression artifacts (\emph{e.g.} missing details of the human body),
		single-frame approaches fail to handle temporal noise, while MFQE 2.0 and STDF-R3 suffer from over-smoothing. However, our method restores much more details or texture than the other methods, especially the details in the boundary areas of fast-moving objects. 
		
		\section{Conclusion}\label{sec:conclusion}
		In this paper, we propose a new method to boost the performance of video compression artifact reduction. Our method consists of two novel modules: the Recursive Fusion~(RF) module and the Deformable Spatiotemporal Attention~(DSTA) module. The former is proposed to capture spatiotemporal information from frames in a large temporal scope, while the latter aims at highlighting the artifact-rich areas in each frame, such as the boundary areas of moving objects. Our extensive experiments demonstrate that our proposed method can achieve superior performance over the state-of-the-art methods. The proposed modules also can be easily adapted to existing multi-frame methods and video-related low-level tasks. 
		
		
		\begin{acks}
		 This work was supported by Zhejiang Lab under grant No.~2019KB0 AB05.
		\end{acks}
		
		\bibliographystyle{ACM-Reference-Format}
		\balance
		\bibliography{main}
		
	\end{document}


	\title{Recursive Fusion and Deformable Spatiotemporal Attention for Video Compression Artifact Reduction \\{\small Supplementary file}}
	
	
	\author{Minyi Zhao$^*$, Yi Xu$^*$, Shuigeng Zhou$^+$}
	\thanks{\textsuperscript{*}Both authors contributed equally to this research.}
	\thanks{\textsuperscript{+}Corresponding author.}
	\affiliation{
		\institution{Shanghai Key Lab of Intelligent Information Processing, and School of Computer Science, Fudan University}
		\city{Shanghai}
		\country{China}
	}
	\email{{zhaomy20,yxu17,sgzhou}@fudan.edu.cn}
	
	
	\maketitle
	
	\begin{table*}
		\centering
		
		\caption{Overall BD-BR reduction ($\%$) of test sequences with the HEVC baseline as an anchor. Calculated at \emph{QP} = 22, 27, 32, 37 and 42.}
		\begin{tabular}{|c|l|| c c c c c c c c c|}
			
			\hline
			\multicolumn{2}{|c||}{\multirow{2}{*}{Sequence}} & AR-CNN & DnCNN &Li \textit{et al.} &DCAD & DS-CNN & {MFQE 1.0} & {MFQE 2.0} & {STDF-R3L}&  Ours\\ [-0.3em]
			
		  \multicolumn{2}{|c||}{}& {\cite{dong2015compression}}  & {\cite{zhang2017beyond}} & {\cite{li2017efficient}} & {\cite{wang2017novel}} & {\cite{yang2018enhancing}} & {\cite{2018Multi}}  & {\cite{2019MFQE}}  & \cite{deng2020spatio} & RFDA \\
			\hline
			\multirow{2}{*}{A} & \textit{Traffic}
			& 7.40 & 8.54 & 10.08 & 9.97 & 9.18 & 14.56 & 16.98 & 21.19  & \textbf{22.70}\\
			
			& \textit{PeopleOnStreet}
			& 6.99 & 8.28 & 9.64 & 9.68 & 8.67 & 13.71 & 15.08  & 17.42  & \textbf{21.11}\\
			
			\cline{1-11}
			\multirow{5}{*}{B} & \textit{Kimono}
			& 6.07 & 7.33 & 8.51 & 8.44 & 7.81 & 12.60 & 13.34  &17.96  & \textbf{22.32}\\
			
			& \textit{ParkScene}
			& 4.47 & 5.04 & 5.35 & 5.68 & 5.42 & 12.04 & 13.66 &18.10& \textbf{19.85}\\
			
			& \textit{Cactus}
			& 6.16 & 6.80 & 8.23 & 8.69 & 8.78 & 12.78 & 14.84  &21.54 & \textbf{21.78} \\
			
			& \textit{BQTerrace}
			& 6.86 & 7.62 & 8.79 & 9.98 & 8.67 & 10.95 & 14.72&\textbf{24.71}  &24.41\\
			
			& \textit{BasketballDrive}
			& 5.83 & 7.33 & 8.61 & 8.94 & 7.89 & 10.54 & 11.85  &16.75 &\textbf{20.24} \\
			
			\cline{1-11}
			\multirow{4}{*}{C} & \textit{RaceHorses}
			& 5.07 & 6.77 & 7.10 & 7.62 & 7.48 &8.83& 9.61  &\textbf{15.62} &14.29\\
			
			& \textit{BQMall}
			& 5.60 & 7.01 & 7.79 & 8.65 & 7.64 & 11.11 & 13.50 &21.12 &\textbf{21.62} \\
			
			& \textit{PartyScene}
			& 1.88 & 4.02 & 3.78 & 4.88 & 4.08 & 6.67 & 11.28 &\textbf{22.24} & 21.11\\
			
			& \textit{BasketballDrill}
			& 4.67 & 8.02 & 8.66 & 9.80 & 8.22 & 10.47 & 12.63  &15.94 &\textbf{18.06} \\
			
			\cline{1-11}
			\multirow{4}{*}{D} & \textit{RaceHorses}
			& 5.61 & 7.22 & 7.68 & 8.16 & 7.35 & 10.41 & 11.55 &15.26 &\textbf{17.57} \\
			
			& \textit{BQSquare}
			& 0.68 & 4.59 & 3.59 & 6.11 & 3.94 & 2.72 & 11.00&\textbf{33.36}  &31.65 \\
			
			& \textit{BlowingBubbles}
			& 3.19 & 5.10 & 5.41 & 6.13 & 5.55 & 10.73 & 15.20 &\textbf{23.54} &22.89\\
			
			& \textit{BasketballPass}
			& 5.11 & 7.03 & 7.78 & 8.35 & 7.49 & 11.70 & 13.43 &18.42 &\textbf{20.42} \\
			
			\cline{1-11}
			\multirow{3}{*}{E} & \textit{FourPeople}
			& 8.42 & 10.12 & 11.46 & 12.21 & 11.13 & 14.89 & 17.50  &\textbf{22.91} & 22.84 \\
			
			& \textit{Johnny}
			& 7.66 & 10.91 & 13.05 & 13.71 & 12.19 & 15.94 & 18.57 &\textbf{24.55} & 23.87 \\
			
			& \textit{KristenAndSara}
			& 8.94 & 10.65 & 12.04 & 12.93 & 11.49 & 15.06 & 18.34 &23.64 &\textbf{24.47} \\
			
			\cline{1-11}
			\multicolumn{2}{|c||}{Average} & 5.59 & 7.36 & 8.20 & 8.89 & 7.85 & 11.41 & 14.06&20.79  & \textbf{21.73} \\
			
			\hline
			
		\end{tabular}
		\label{tab-bdr}
	\end{table*}

	Tab.~\ref{tab-bdr} shows the results of BD-BR reduction of test sequences with the HEVC baseline as an anchor. All BD-BR are calculated with \textit{QP}={22, 27, 32, 37, 42}, except that STDF~\cite{deng2020spatio} is done with four \textit{QPs}.

	
	\bibliographystyle{ACM-Reference-Format}
	\bibliography{main}